\documentclass[twocolumn,trackchanges]{aastex7}

\begin{document}

\title{SN~2024abvb: A Type Icn Supernova in the Outskirts of its Host Galaxy}

\author[0000-0003-3031-6105]{Maokai Hu 
}
\affiliation{Physics Department, Tsinghua University, Beijing 100084, China}
\email[show]{kaihukaihu123@mail.tsinghua.edu.cn \\}

\author[0009-0004-4256-1209]{Shengyu Yan}
\affiliation{Physics Department, Tsinghua University, Beijing 100084, China}
\email{yansy19@mails.tsinghua.edu.cn}

\author[0000-0002-7334-2357]{Xiaofeng Wang}
\affiliation{Physics Department, Tsinghua University, Beijing 100084, China} 
\email[show]{wang\_xf@mail.tsinghua.edu.cn}

\author[0009-0003-9229-9942]{Abdusamatjan Iskandar}
\affiliation{Xinjiang Astronomical Observatory, Chinese Academy of Sciences, Urumqi, Xinjiang, 830011, China}
\affiliation{School of Astronomy and Space Science, University of Chinese Academy of Sciences, Beijing 100049,  China}
\email{abudu@xao.ac.cn}

\author[0000-0002-8296-2590]{Jujia Zhang}
\affiliation{Yunnan Observatories, Chinese Academy of Sciences, Kunming 650216, China}
\affiliation{International Centre of Supernovae, Yunnan Key Laboratory, Kunming 650216, China}
\affiliation{Key Laboratory for the Structure and Evolution of Celestial Objects, Chinese Academy of Sciences, Kunming 650216, China}
\email{jujia@ynao.ac.cn}

\author[]{Liping Li}
\affiliation{Yunnan Observatories, Chinese Academy of Sciences, Kunming 650216, China}
\affiliation{International Centre of Supernovae, Yunnan Key Laboratory, Kunming 650216, China}
\affiliation{Key Laboratory for the Structure and Evolution of Celestial Objects, Chinese Academy of Sciences, Kunming 650216, China}
\email{liliping@ynao.ac.cn}

\author{Ali Esamdin} 
\affiliation{Xinjiang Astronomical Observatory, Chinese Academy of Sciences, Urumqi, Xinjiang, 830011, China}
\affiliation{School of Astronomy and Space Science, University of Chinese Academy of Sciences, Beijing 100049,  China}
\email{aliyi@xao.ac.cn}

\author{Letian Wang} 
\affiliation{Xinjiang Astronomical Observatory, Chinese Academy of Sciences, Urumqi, Xinjiang, 830011, China}
\email{wangletian@xao.ac.cn}

\author[0000-0002-1094-3817]{Lingzhi Wang}
\affiliation{College of Science, Hainan Tropical Ocean University, Sanya, 572022, China}
\affiliation{Chinese Academy of Sciences South America Center for Astronomy (CASSACA), National Astronomical Observatories, CAS, Beijing, China}
\email{wanglingzhibnu@gmail.com}

\author[0000-0003-3460-0103]{Alexei V. Filippenko}  
\affiliation{Department of Astronomy, University of California, Berkeley, CA 94720-3411, USA}
\affiliation{Hagler Institute for Advanced Study, Texas A\&M University, 3572 TAMU, College Station, TX 77843, USA}
\email{afilippenko@berkeley.edu}  

\author[0000-0001-5955-2502]{Thomas G. Brink}  
\affiliation{Department of Astronomy, University of California, Berkeley, CA 94720-3411, USA}
\email{tgbrink@berkeley.edu} 


\author[]{Liyang Chen}
\affiliation{Physics Department, Tsinghua University, Beijing 100084, China}
\email{chenly23@mails.tsinghua.edu.cn}

\author[]{Ruifeng Huang}
\affiliation{Physics Department, Tsinghua University, Beijing 100084, China}
\email{huangrf24@mails.tsinghua.edu.cn}

\author[0000-0001-7092-9374]{Lifan Wang}
\affiliation{George P. and Cynthia Woods Mitchell Institute for Fundamental Physics \& Astronomy, Texas A. \& M. University, 4242 TAMU, College Station, TX 77843, USA}
\email{lifan@tamu.edu}


\begin{abstract}

We present multiband photometric and spectroscopic observations of supernova (SN)~2024abvb, which exhibits early-time prominent photoionized narrow emission lines of C{\sc\,II} superposed on a blue continuum. The absence of Balmer features indicates that the SN exploded within hydrogen-poor circumstellar matter (CSM). Together with the lack of explicit evidence of helium signatures, we tentatively identify SN~2024abvb as a Type Icn SN (SN~Icn). After correcting for extinction, we estimate an $r$-band peak absolute magnitude of $-19.7\pm 0.1$, placing SN~2024abvb in the luminous regime of SNe~Icn. We adopted a hybrid model that accounts for both the energy released by the ejecta-CSM interaction and the radioactive decay of nickel synthesized in the SN ejecta to fit the light curve of SN~2024abvb. The best-fit model to the multiband light curves within the first $\sim 40$\,days after explosion suggests that the CSM, radioactive nickel, and ejecta masses to be $0.28^{+0.02}_{-0.03}\,M_{\odot}$, $\le3.8\times10^{-2}$\,$M_{\odot}$, and $0.12^{+0.06}_{-0.02}\,M_{\odot}$, respectively. Such a low ejecta mass indicates that the progenitor star of SN~2024abvb experienced a significant mass-stripping process, consistent with the hydrogen-poor and helium-poor spectral features. SN~2024abvb provides important insights into the physical origins of the rare subclass of SNe~Icn.

\end{abstract}



\section{Introduction} 

Interacting supernovae (SNe) are a unique type of transient characterized by strong interaction between the SN ejecta and circumstellar matter (CSM; \citealt{2017hsn..book..403S}). The presence of the latter may obscure the prompt emission from the former to some extent, depending on the details of the ejecta kinematics and the CSM distribution. The resulting spectral signatures manifest as a series of narrow photoionized emission lines superposed on a blue continuum (e.g., \citealt{2001MNRAS.325..907F, 2004MNRAS.352.1213C, 2010ApJ...709..856S, 2012AJ....144..131Z,2014Natur.509..471G, 2016ApJ...818....3K, 2017NatPh..13..510Y, 2017A&A...605A..83D, 2021ApJ...912...46B, 2023ApJ...952..119B, 2024ApJ...970..189J}). Substantial efforts have gone into establishing the relationship between the physical properties of the surrounding CSM and the mass-loss history of progenitor stars, challenging the existing paradigms of stellar evolution. 

One major hydrogen-deficient type of interacting events involves Type Ibn SNe (SNe~Ibn), with a fraction of about 8.8\% among the Type Ibc sample\citep{2025A&A...698A.305M}. Such transients are characterized by narrow helium emission lines in their spectra, indicating the presence of He-rich and H-poor CSM with an expansion velocity of $\sim 1000\,{\rm km}\,{\rm s}^{-1}$ \citep{2008MNRAS.389..113P,2015MNRAS.449.1921P,2015MNRAS.449.1954P,2016MNRAS.456..853P,2017MNRAS.471.4381S,2020ApJ...889..170G,d2025arXiv250610700G}. The chemical composition and the expansion velocity of the CSM lead to the interpretation that SNe~Ibn arise from the terminal explosion of Wolf-Rayet (WR) stars \citep{2007ApJ...657L.105F,2008MNRAS.389..113P,2017ApJ...836..158H,2022A&A...658A.130D,2022ApJ...927...25M}. In particular, the pre-explosion outburst of SN~2006jc provided strong evidence for transitional objects between luminous blue variables and WR stars as the progenitors of SNe~Ibn \citep{2007Natur.447..829P}. However, several issues remain unresolved to establish the connection between SNe~Ibn and their possible progenitor WR stars. For instance, there is a discrepancy between the observed rate of SNe~Ibn and the estimated explosion rate of SNe~Ibn based on the Galactic WR stars, as well as between the inferred mass-loss rate of the CSM around SNe~Ibn and WR stars \citep{2007ARA&A..45..177C,2014ARA&A..52..487S}.

A more exotic category of interacting SNe, denoted as ``Icn,'' has recently been identified. The discovery event, SN~2019hgp, exhibits distinctive narrow emission lines of carbon and oxygen but lacks signatures of H and He \citep{2022Natur.601..201G}. Based on a detailed investigation of its circumstellar environment, \cite{2022Natur.601..201G} suggested that SN~2019hgp arises from the explosion of a WR star with a helium-poor and carbon-rich stellar wind. Up to now, only four events have been reported as SNe~Icn (SNe~2019jc, 2021ckj, 2021csp, 2022ann; \citealt{2021arXiv210807278F,2022ApJ...927..180P,2022ApJ...938...73P,2023MNRAS.523.2530D,2023A&A...673A..27N}). Despite the scarcity of high-quality data sets, SNe~Icn present diverse properties in their light curves and host-galaxy environments, indicating multiple progenitor channels \citep{2022ApJ...938...73P}. In particular, SN~2019jc exhibits a relatively low peak luminosity and explosion energy \citep{2022ApJ...938...73P}, which may indicate alternative progenitor channels, such as the merger of an oxygen-neon white dwarf and a carbon-oxygen white dwarf~\citep{2024ApJ...967L..45W}.

SN~2024abvb was discovered on UTC Nov. 22.33, 2024 (MJD\,60636.334; \citealt{2024TNSTR4579....1T}) by the Asteroid Terrestrial-impact Last Alert System (ATLAS; \citealt{2018PASP..130f4505T}). Follow-up spectroscopy obtained on MJD\,60641.897 revealed a series of photoionized narrow C{\sc\,II} emission lines superposed on a blue continuum, thereby putting it into the rare class of SNe~Icn \citep{2024TNSCR4674....1S}. In Sec.~\ref{SecII}, we describe our multiband photometric and spectroscopic observations of SN~2024abvb and the data reduction. Sec.~\ref{SecIII} presents a more detailed analysis of the light curve and spectra. Our discussion and conclusions are given in Sec.~\ref{SecIV}.

\begin{figure*}[ht!]
\plotone{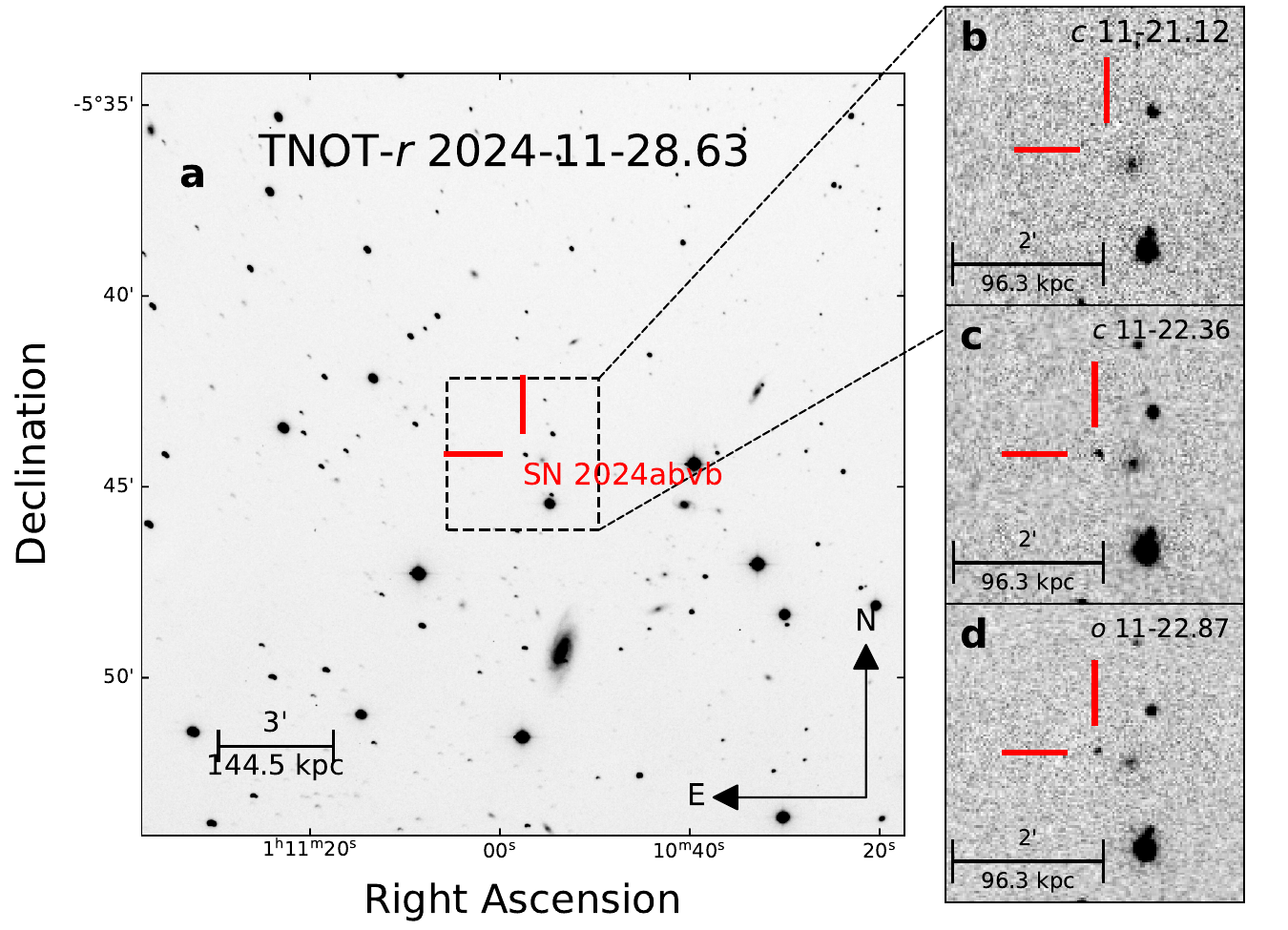}
\caption{Panel (a): TNOT $r$-band image showing the location of SN~2024abvb. The black dashed square outlines the sky regions displayed in panels (b), (c), and (d), which show respectively the latest $c$-band prediscovery image, and the $c$- and $o$-band first-detection images obtained by ATLAS. Epochs are labeled in the upper right of each panel. 
\label{fig_11_finder}}
\end{figure*}

\begin{figure*}[ht!]
\plotone{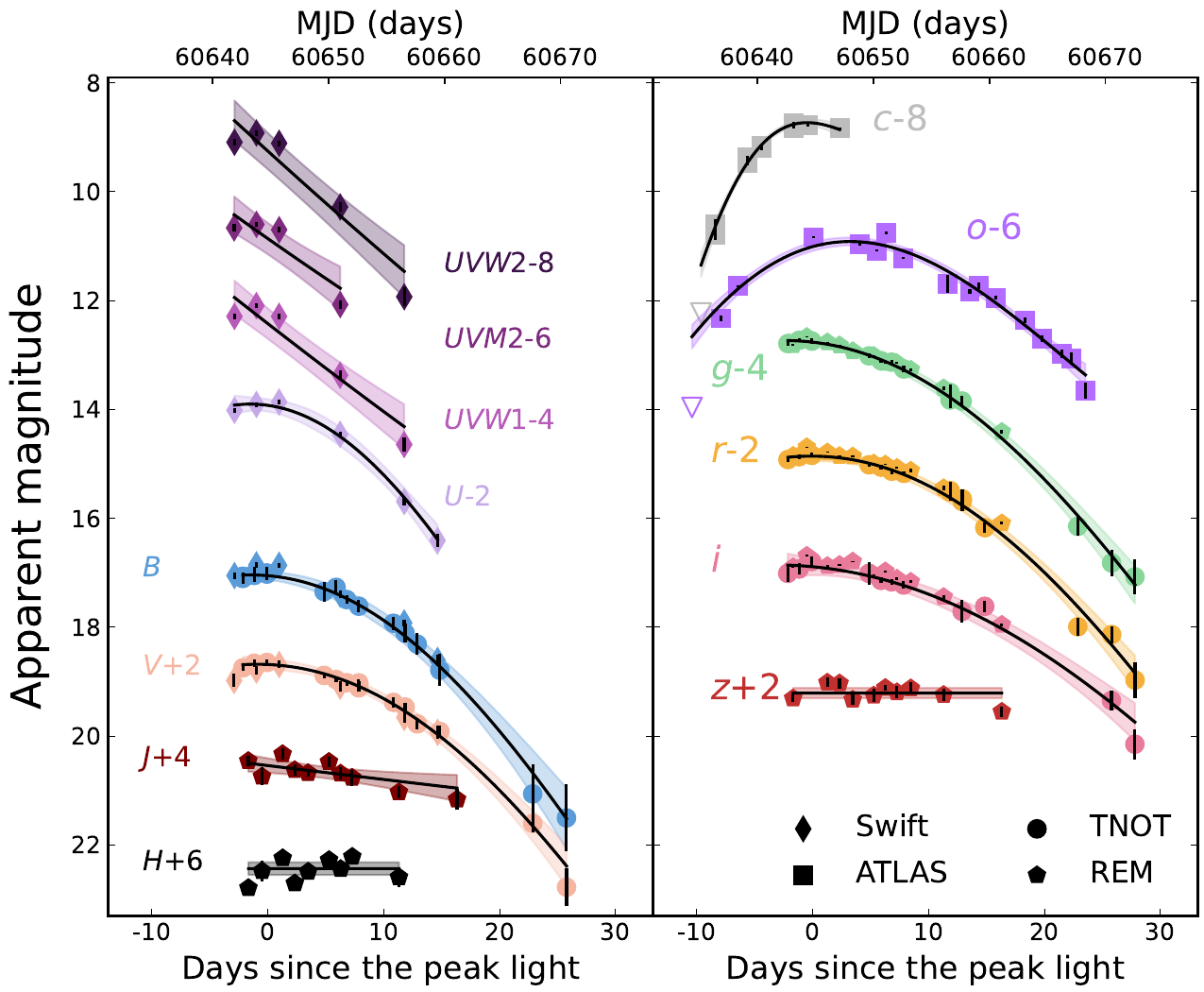}
\caption{Multiband light curves of SN~2024abvb. The bandpass of each light curve is marked next to the color-coded symbols, and instruments used to obtain the photometry are indicated by the legend. The open upside-down triangles show the latest nondetection limits in the ATLAS $c$ and $o$ bands. The black curve and the color-shaded region underlying the photometry for each band provide the smoothed light curve and its associated uncertainty, respectively, obtained via a Gaussian process. The phase is relative to the $r$-band peak light of SN~2024abvb.
\label{fig_11_LCs}}
\end{figure*}

\section{Observations} 
\label{SecII}

We measure a radial velocity of $-11,700$\,km\,s$^{-1}$ of SN~2024abvb by matching the centroids of a series of narrow C{\sc\,II} emission lines (e.g., C{\sc\,II}$\lambda$5890, C{\sc\,II} $\lambda$6580, and C{\sc\,II}$\lambda$7235) in the classification spectrum \citep{2024TNSCR4674....1S}. We interpret the radial velocity as exclusively due to a redshift $z = 0.039$ and adopt a Hubble constant of H$_{0}=70$\,km\,s$^{-1}$\,Mpc$^{-1}$. We therefore derive a distance to SN\,2024abvb of 165\,Mpc assuming a standard $\Lambda$CDM cosmology. 

Adopting the position reported by the ATLAS discovery stream (RA $= 01^{\rm hr}10^{\rm m}57.550^{\rm s}$, Dec. $= -05^\circ44'08.06''$; J2000) and assuming an $R_{V}=3.1$ extinction law as the Galactic average~\citep{1989ApJ...345..245C},  we estimate a Galactic reddening of $E(B-V)_{\rm mw} = 0.179\,{\rm mag}$ along the line of sight to SN\,2024abvb by querying the NASA/IPAC NED Galactic Extinction calculator\footnote{\url{https://ned.ipac.caltech.edu/forms/calculator.html}} and the extinction map by \citet{2011ApJ...737..103S}. We identify no obvious Na\,{\sc\,I}\,D\,$\lambda\lambda$5890, 5896 absorption doublet at the inferred redshift of the SN, whose line strength has been proposed as a tracer of the amount of the interstellar extinction~\citep{1997A&A...318..269M,2011MNRAS.415L..81P}. As shown in Figure~\ref{fig_11_finder}, we note that SN\,2024abvb exploded in the outskirts of a galaxy whose centroid is measured as RA $= 01^{\rm hr}10^{\rm m}55.7^{\rm s}$, Dec. $= -05^\circ44'17''$ (J2000), which yields a rather large separation between the SN and its tentative host ($\sim 29''$; and 23\,kpc). We suggest that the line-of-sight reddening toward SN\,2024abvb is dominated by the Galactic component and made no attempt to derive the extinction from its host.

\subsection{Photometry}    

We present multiband light curves of SN~2024abvb in Figure~\ref{fig_11_LCs} and list all the photometry in Table~\ref{Table_33_LCs}. 

\subsubsection{TNOT}

The $BVgri$-band light curves of SN~2024abvb were obtained with the Tsinghua-Nanshan Optical Telescope (TNOT), an 80\,cm reflector that began operating in 2024. TNOT is an equatorial telescope located at Nanshan Station of Xinjiang Astronomy Observatories, Chinese Academy of Sciences. The telescope system features an ASA800 Ritchey-Chr\'etien optical tube ($f$/6.85, 5480\,mm focal length) on ASA's premium DDM500 equatorial mount. The field of view is designed to be $26.5' \times 26.5'$. The Bessel $BV$ system and the Sloan $ugri$ system are chosen as the primary filter set. The imaging system utilizes an Andor iKon-L DZ936 CCD camera with a $2048 \times 2048$ pixel array, and the plate scale is 0\farcs{78}\,pixel$^{-1}$.

All images obtained by TNOT were preprocessed using standard \texttt{IRAF} \footnote{{IRAF} 
is distributed by the National Optical Astronomy Observatories, which are 
operated by the Association of Universities for Research in Astronomy, Inc., 
under cooperative agreement with the U.S. National Science Foundation (NSF).} 
routines~\citep{1986SPIE..627..733T, 1993ASPC...52..173T}, including bias subtraction and flat-field correction. For each epoch of the TNOT observation, 3--5 exposures were taken per filter. After computing the World Coordinate System solutions, we employed the \texttt{reproject} package\footnote{\href{https://reproject.readthedocs.io/en/stable/index.html}{https://reproject.readthedocs.io/en/stable/index.html}} to align and combine the individual images. Point-spread-function (PSF) photometry was then performed using \texttt{AutoPhOT} \citep{brennanAUTOmatedPhotometryTransients2022}. 
Instrumental magnitudes of the SN were extracted using the \texttt{AutoPhOT} \citep{brennanAUTOmatedPhotometryTransients2022} pipeline. In particular, for each frame, the PSF was constructed from bright, isolated field stars and matched to the SN. The fitting radius of the PSF model was set to the full width at half-maximum intensity (FWHM), and the background was estimated and subtracted iteratively during the PSF fit. For photometric calibration, we adopted the Pan-STARRS catalog for the $g$, $r$, and $i$ bands, and the APASS catalog for the Johnson $B$ and $V$ bands. Because the SN is bright and in the outskirts of its host galaxy, background removal by means of subtracting archival template images is not necessary.


\subsubsection{ATLAS}

We also include the $c$- and $o$-band data of SN~2024abvb from ATLAS using forced photometry\footnote{\href{https://fallingstar-data.com/forcedphot/}{https://fallingstar-data.com/forcedphot/}} \citep{2018PASP..130f4505T,2020PASP..132h5002S}. Upper limits were derived based on a $5\sigma$ threshold above the flux uncertainties. Since the ALTAS footprint covered the field of SN\,2024abvb $\sim3$--4 times per night, we stacked the nightly photometry to improve the signal-to-noise ratio (S/N).

\subsubsection{Swift}
The UltraViolet/Optical Telescope (UVOT) onboard the Neil Gehrels \textit{Swift} Observatory conducted the first set of observations of SN\,2024abvb on UTC~2024 Nov. 27 at 20:22:27.50 (MJD = 60641.85) in the $UVW1$, $UVM2$, $UVW2$, and optical $ubv$ bands (\texttt{ObsID 00018944}) \citep{2004ApJ...611.1005G,2005SSRv..120...95R}, spanning from $\sim -2.9$ to $\sim +35.1$ days corresponding to the $r$-band peak brightness. We performed aperture photometry using a Python-based pipeline that wraps around \texttt{HEAsoft} tools such as \texttt{uvotmaghist} and \texttt{uvotimsum} \citep{2009AJ....137.4517B,2014Ap&SS.354...89B}.\footnote{The home page of the code on GitHub: \url{https://github.com/gterreran/Swift_host_subtraction}} A 5$^{\prime\prime}$-radius aperture on the target and a 20$^{\prime\prime}$-radius aperture on the background are adopted for the measurements. Host-galaxy contamination at the SN location is minimal, so host subtraction was not applied to the photometry. 

\subsubsection{REM}

The $grizJH$-band light curves of SN~2024abvb were obtained with the Rapid Eye Mount (REM) telescope \citep{Covino:2004SPIE.5492.1613C,Zerbi:2004SPIE.5492.1590Z}, which is a fully automated, 60\,cm fast-response Ritchey-Chr\'etien' telescope operating at the ESO La Silla Observatory in Chile. REM is equipped with two instruments: REMIR, a near-infrared (NIR) imaging camera \citep{Conconi:2004SPIE.5492.1602C}, and ROSS2, a visible-light camera \citep{Tosti:2004SPIE.5492..689T}. Both cameras simultaneously cover nearly the same $10' \times 10'$  field of view. With its dichroic beam-splitting design, REM can capture five images at once. 
Astrometric calibration of the optical and NIR images was performed using Astrometry.net \citep{Lang:2010AJ....139.1782L}. For each band, if more than two images with identical exposure times were available, they were combined using SWarp \citep{swarp:2010ascl.soft10068B} to produce a deeper stacked image. Aperture photometry was then carried out on all images using SExtractor \citep{sextracotr:1996A&AS..117..393B}. Owing to the small field of view, photometric calibration was performed using an isolated bright reference star from the ATLAS-REFCAT2 catalog \citep{Tonry:2018ApJ...867..105T}.

\begin{figure*}[ht!]
\plotone{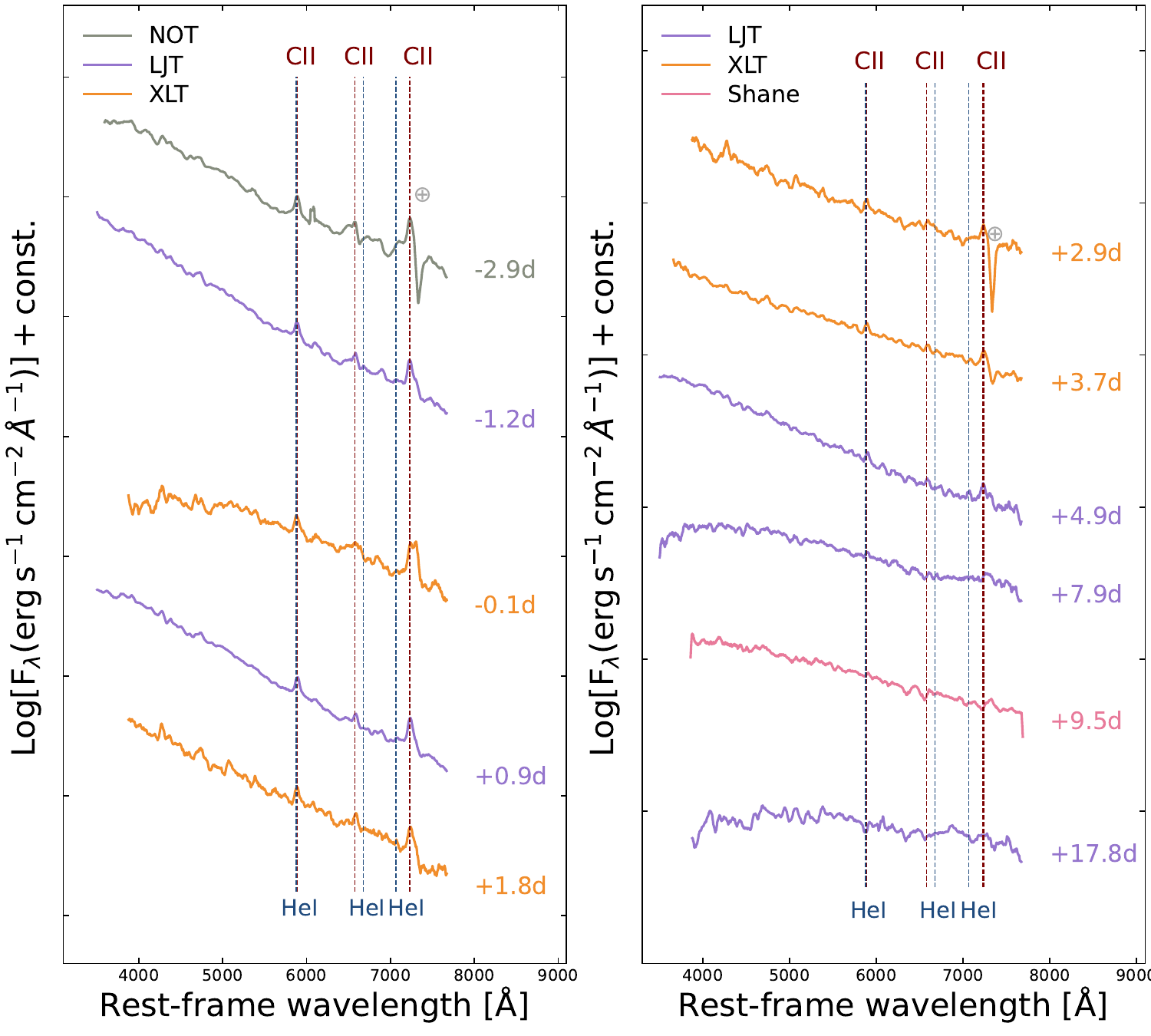}
\caption{Spectral time series of SN\,2024abvb acquired by NOT, LJT, XLT, and Lick 3\,m Shane from approximately days $-3$ to $+$18 relative to the time of $r$-band maximum brightness (see Sec~\ref{Sec3.1}). All data are presented with 50\,\AA\ binning. The gray, purple, orange, and pink curves show the spectra obtained by NOT, LJT, XLT, and Shane, respectively. The vertical dashed lines indicate the spectral lines of C{\sc\,II} and He{\sc\,I}. The machine-readable table of the listed spectra is available online.
\label{fig_22_spe}}
\end{figure*}

\subsection{Spectroscopy}

The spectral time sequence of SN~2024abvb was obtained by the Xinglong 2.16\,m telescope (XLT; \citealt{2016PASP..128k5005F}), the Lijiang 2.4\,m telescope (LJT; \citealt{2019RAA....19..149W}), and the Shane 3\,m telescope. Spectral reductions were performed using standard procedures within the \texttt{IRAF} software package, including steps such as bias subtraction, flat-field normalization, and cosmic-ray removal. Wavelength calibration was carried out using comparison-lamp exposures. Flux calibration was implemented using nightly exposures of spectrophotometric standard stars at air masses similar to those of SN\,2024abvb. Telluric lines were removed by using an average telluric spectrum constructed from multiple observations of telluric standard stars scaled to match the ﬂux level of the SN spectra. For comparison, we also present the classification spectrum obtained by the Nordic Optical Telescope (NOT) \citep{2024TNSCR4674....1S}. 

The complete spectral sequence of SN\,2024abvb discussed in this paper is shown in Figure~\ref{fig_22_spe}. A log of spectroscopic observations is presented in Table~\ref{Table_22_SpecInfor}.

\begin{figure*}[ht!]
\plotone{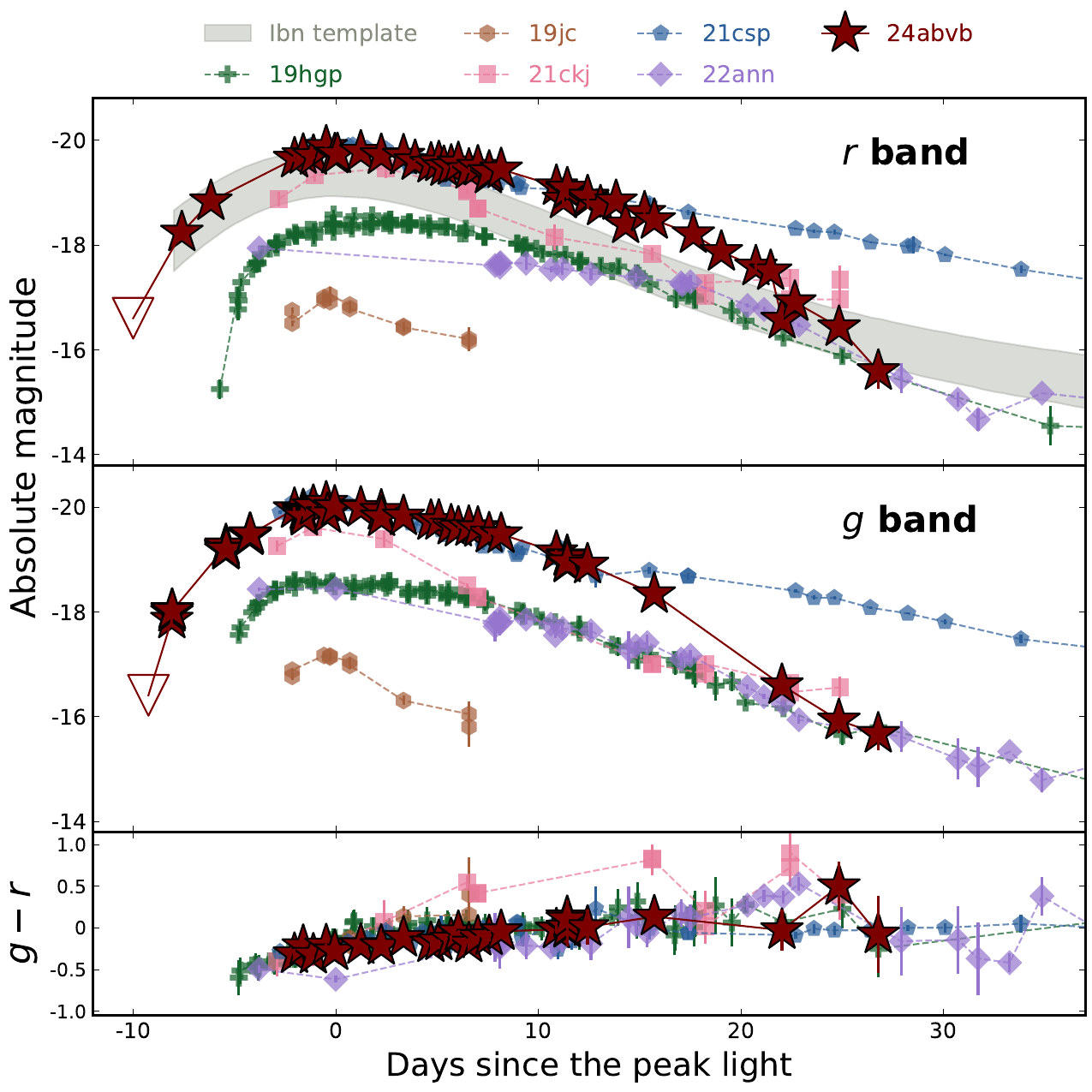}
\caption{{\it Upper and middle panels:} the $r$- and $g$-band light curves of SN\,2024abvb compared to those of other well-sampled SNe~Icn, including SNe\,2019hgp~\citep{2022Natur.601..201G}, 2019jc~\citep{2022ApJ...938...73P}, 2021ckj~\citep{2022ApJ...938...73P, 2023A&A...673A..27N}, 2021csp~\citep{2022ApJ...927..180P, 2022ApJ...938...73P}, and 2022ann~\citep{2023MNRAS.523.2530D}.
The gray-shaded area presents the template light curve derived by averaging a sample of SNe~Ibn \citep{2017ApJ...836..158H}. 
{\it Lower panel:} comparison of the $g-r$ color of SN\,2024abvb and other SNe~Icn. The phase of SN\,2024abvb is relative to its $r$-band maximum light, and other objects are from their references. Note that the light curve of each SN has been corrected for reddening, and all phases have been corrected for time dilation caused by the redshift.  
\label{fig_33_LCs_Icn}}
\end{figure*} 

\section{Analysis}
\label{SecIII}

\subsection{Photometric Properties}
\label{Sec3.1}

As shown in Figure~\ref{fig_11_LCs}, we smoothed the multiband light curves of SN~2024abvb with a Gaussian process and estimated that the peak of the $r$-band light curve is at MJD $= 60644.75$. Throughout the paper, all phases are given relative to the time of the $r$-band peak brightness at MJD\,60644.75. The last nondetection limit and the first detection in the ATLAS $o$ band yield $19.96$\,mag at day $-$10.4 and $18.33\pm0.06$\,mag at day $-$7.9, respectively. The former has been assigned to constrain the explosion time when fitting the multiband early light curves of SN\,2024abvb (see  Sec.~\ref{Model_lc}).

The $r$-band light curve of SN\,2024abvb reached a maximum of $16.9\pm0.1$\,mag at MJD\,60644.75, corresponding to an absolute magnitude of $-19.7\pm0.1$\,mag with the Galactic extinction correction. The $r$-band peak brightness of SN\,2024abvb is comparable to that of SN\,2021csp~\citep{2022ApJ...927..180P}, both of which represent cases at the luminous end of the SN~Icn distribution, as illustrated in the upper panel of Figure~\ref{fig_33_LCs_Icn}. Notably, SN\,2024abvb exhibits a rapid drop after $\sim 10$ days compared to SN\,2021csp. Except for an overall $\sim 0.5$\,mag excess in brightness, the light-curve morphology of SN~2024abvb more resembles that of SN~2021ckj. The latter overlaps with the template light curve of SNe~Ibn as presented by the gray-shaded band in the upper panel of Figure~\ref{fig_33_LCs_Icn}.

In the bottom panel of Figure~\ref{fig_33_LCs_Icn} we present the $g-r$ color evolution of SN\,2024abvb spanning $\sim 5$ days before to 37 days after the $r$-band light-curve peak. A monotonic change in $g-r$ color from about $-0.25$ to 0.0\,mag can be identified in the time interval $-5$ to 16 days during which the multiband light curves are well sampled.

\begin{figure*}[ht!]
\plotone{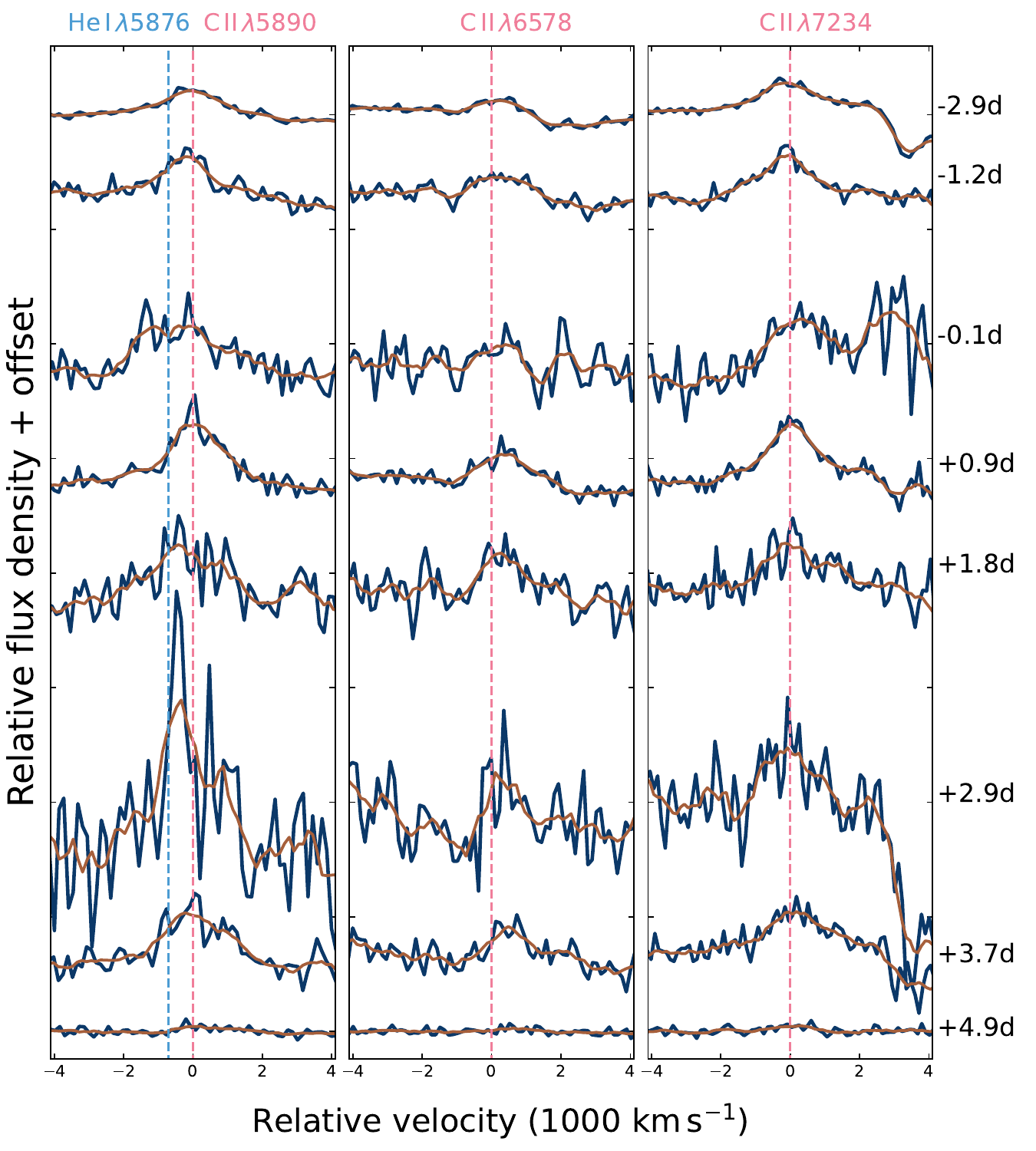}
\caption{Evolution of C\,{\sc\,II} emission lines of SN~2024abvb in velocity space relative to their rest-frame wavelengths. The dark-blue lines represent the arbitrarily scaled flux spectra in full resolution. The brown curves display the smoothed spectra. Phases are labeled on the right. Vertical pink lines mark the rest-frame wavelengths of the C{\sc\,II} lines of interest. For illustration, the vertical dashed blue line in the left panel indicates the rest-frame wavelength of He{\sc\,I} $\lambda5876$.
\label{fig_77_spe_velocity}}
\end{figure*}

\begin{figure*}[ht!]
\plotone{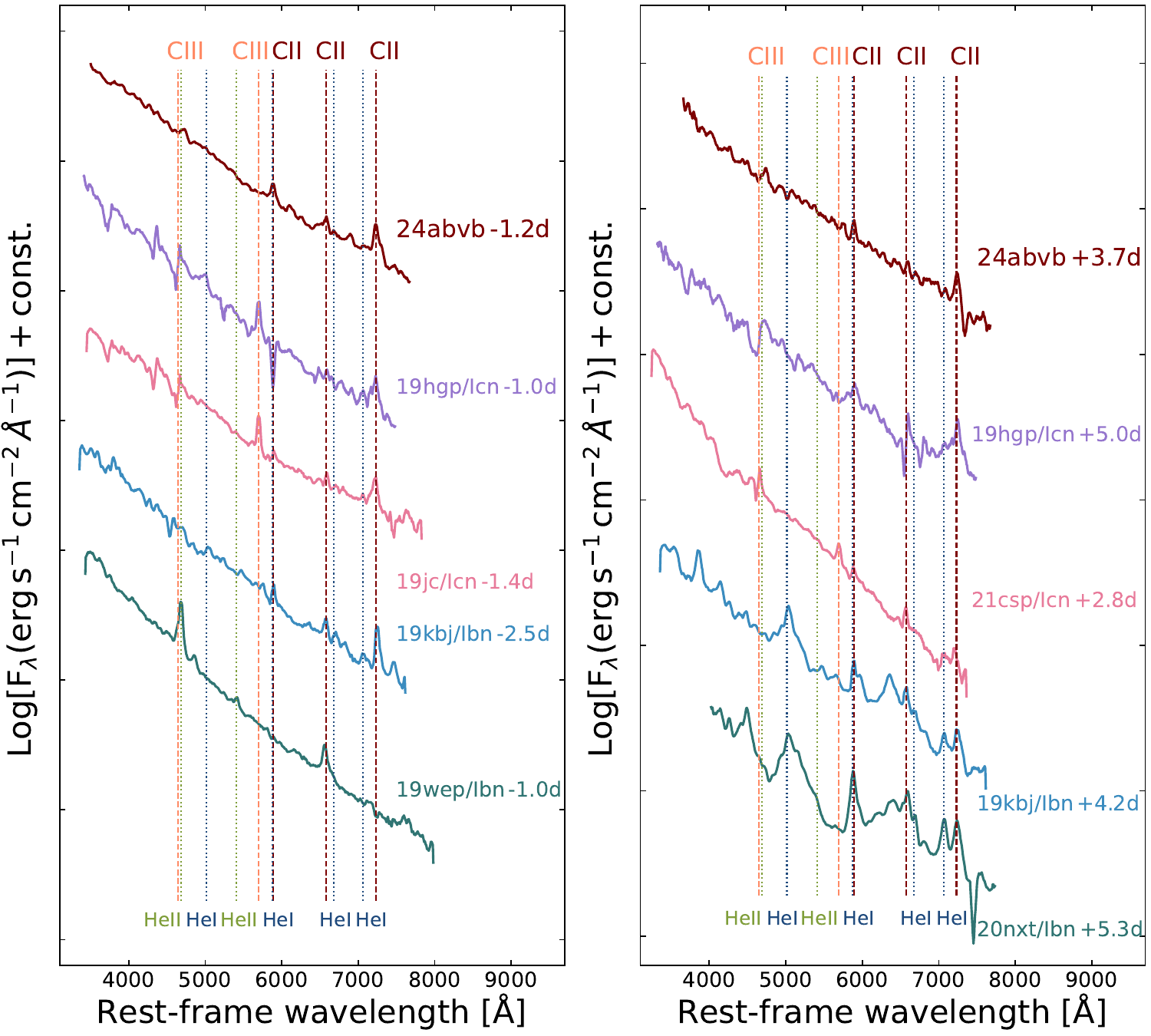}
\caption{The day $-1.2$ (left-hand panel) and $+3.7$ (right-hand panel) spectra of SN~2024abvb compared with spectra of a selected sample of well-observed SNe~Ibn/Icn at similar phases, namely SNe~2019kbj, 2019wep, 2020nxt (Ibn; \citealt{2022ApJ...930..127G,2023ApJ...946...30B,2024MNRAS.530.3906W}) and SNe~2019hgp, 2019jc, 2021csp (Icn; \citealt{2022ApJ...938...73P,2022ApJ...927..180P}). The vertical dashed lines mark the rest-frame wavelengths of prominent carbon features, and the vertical dotted lines indicate the positions of several major helium lines. All spectra are presented after being binned by a $\sim 50$-\AA\ kernel.
\label{fig_88_spe_compa}}
\end{figure*}

\subsection{Spectroscopic Properties}

The early-time spectra of SN~2024abvb are characterized by three emission features at rest-frame wavelengths near 5890\,\AA, 6580\,\AA, and 7235\,\AA\ superimposed on a blue continuum as shown in Figure~\ref{fig_22_spe}. Portraits of photoionized spectral features centered at several C{\sc\,II} features, whose rest-frame wavelengths are 5890\,\AA, 6578\,\AA, and 7234\,\AA, are shown respectively in the left, middle, and right columns of Figure~\ref{fig_77_spe_velocity}. These photonionization-powered narrow emission lines, whose FWHM are found to be $\lesssim(1-2)\times10^{3}$\,km\,s$^{-1}$, have vanished by day $+4.9$. Limited by the insufficient S/N, we do not attempt to investigate the temporal evolution of the width of these short-lived emission lines. Despite the proximity to He{\sc\,I}\,$\lambda$5876, owing to the lack of other features of  He{\sc\,I} such as near 6678 and 7065\,\AA, we attribute the emission profile presented in the left column of Figure~\ref{fig_77_spe_velocity} to C{\sc\,II}\,$\lambda$5890 and conclude the absence of helium in the early-phase spectra of SN\,2024abvb.

Figure~\ref{fig_88_spe_compa} compares the spectra of SN\,2024abvb at days $-$1.2 and $+$3.7 with the spectra of a selected sample of well-observed SNe~Ibn/Icn at similar phases. The spectral features of SN\,2024abvb exhibit a strong resemblance to those of SNe~Icn, in which the C{\sc\,II} emission lines are prominent and helium lines are absent. We also note that, unlike other SNe~Icn, the early-time spectra of SN\,2024abvb display no sign of any C{\sc\,III} feature. A natural explanation for this discrepancy could be a relatively lower temperature in the photoionized CSM of SN\,2024abvb, where the highest temperature produces the highest excitation states.

\begin{figure*}[ht!]
\plotone{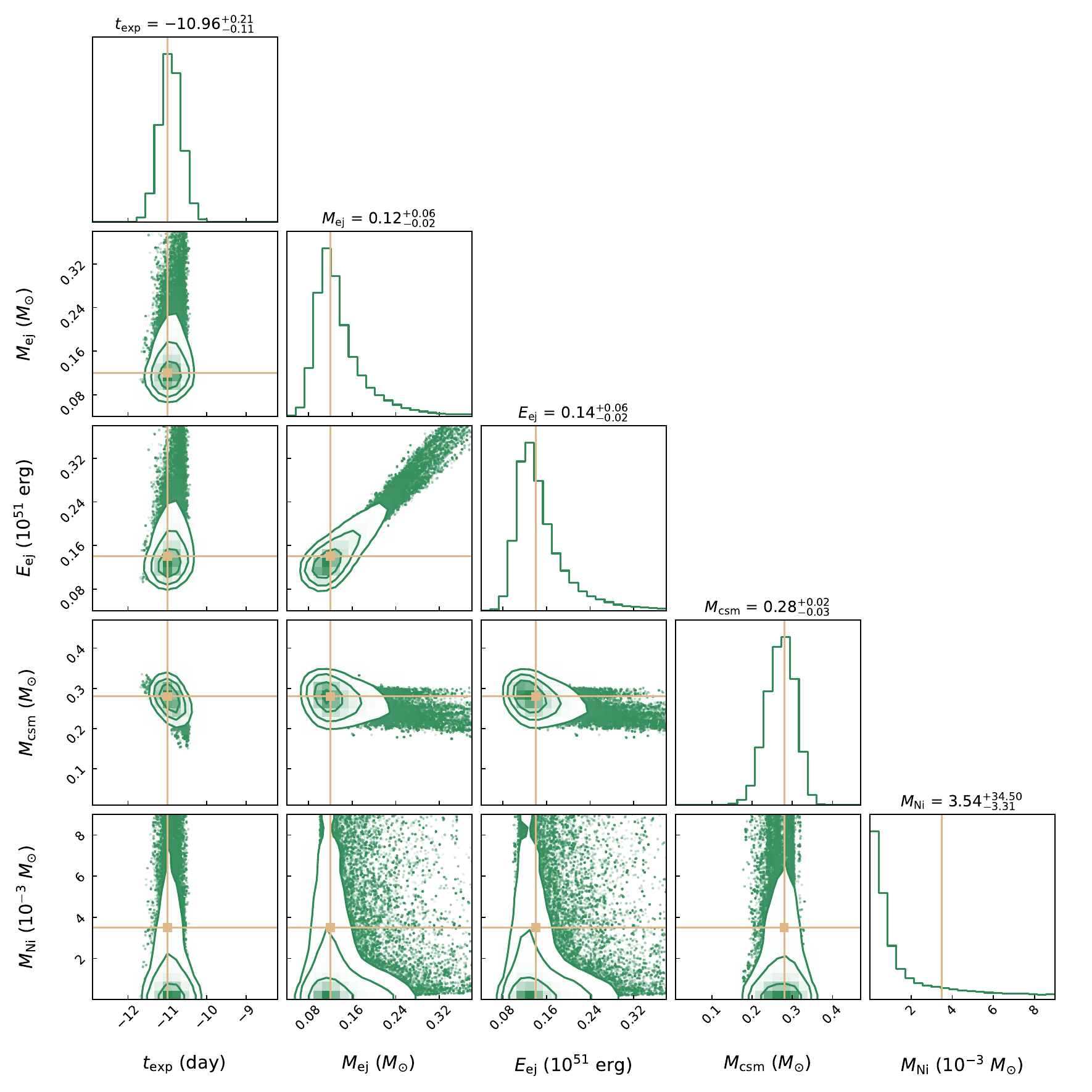}
\caption{Posterior distributions of the hybrid model that includes the CSM interaction and the $^{56}$Ni$\rightarrow ^{56}$Co$\rightarrow ^{56}$Fe decay from the MOSFiT code. Median values are marked by horizontal and vertical orange lines and labeled together with their 1$\sigma$ uncertainties. These are used as the best-fit values. The explosion time $t_{\rm exp}$ is given relative to the time of the $r$-band light-curve peak.
\label{fig_55_mcmc}}
\end{figure*}

\begin{figure*}[ht!]
\plotone{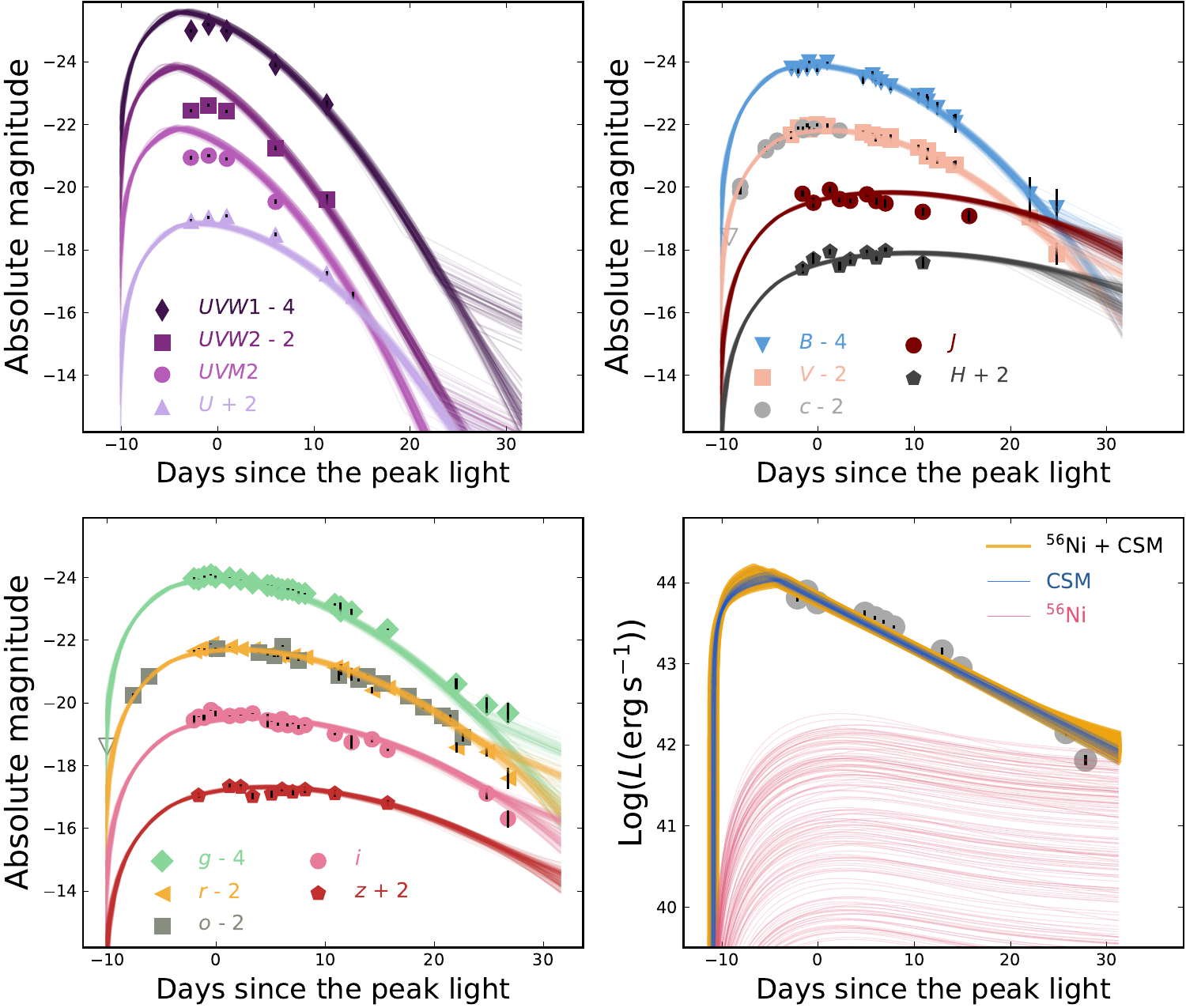}
\caption{The best fit to the multiband light curves of SN~2024abvb with a hybrid model, which accounts for the CSM interaction and the $^{56}$Ni decay. Each color-coded band is formed by stacking 100 randomly selected models generated from MOSFiT. The upper-left, upper-right, and lower-left panels present the light curves from observations in {\it Swift} filters, Bessell $BVJH$, and SDSS $griz$ bandpasses, respectively. The $c$- and $o$-band ATLAS observations were used as $V$- and $r$-band light curves in the fitting procedure. The lower-right panel displays the corresponding comparison for the quasibolometric light curve of SN~2024abvb. The red, blue, and yellow lines represent the components of the $^{56}$Ni decay, the ejecta-CSM interaction, and a combination of these two, respectively. The gray circles in the lower-right panel are the bolometric luminosity of SN~2024abvb calculated by MOSFiT. The yellow lines overlap with the blue lines, as the CSM interaction dominates over radiation near peak brightness. All phases are aligned with the $r$-band maximum and have been corrected for time dilation due to the redshift.
\label{fig_66_LCsFit}}
\end{figure*}

\begin{figure*}[ht!]
\plotone{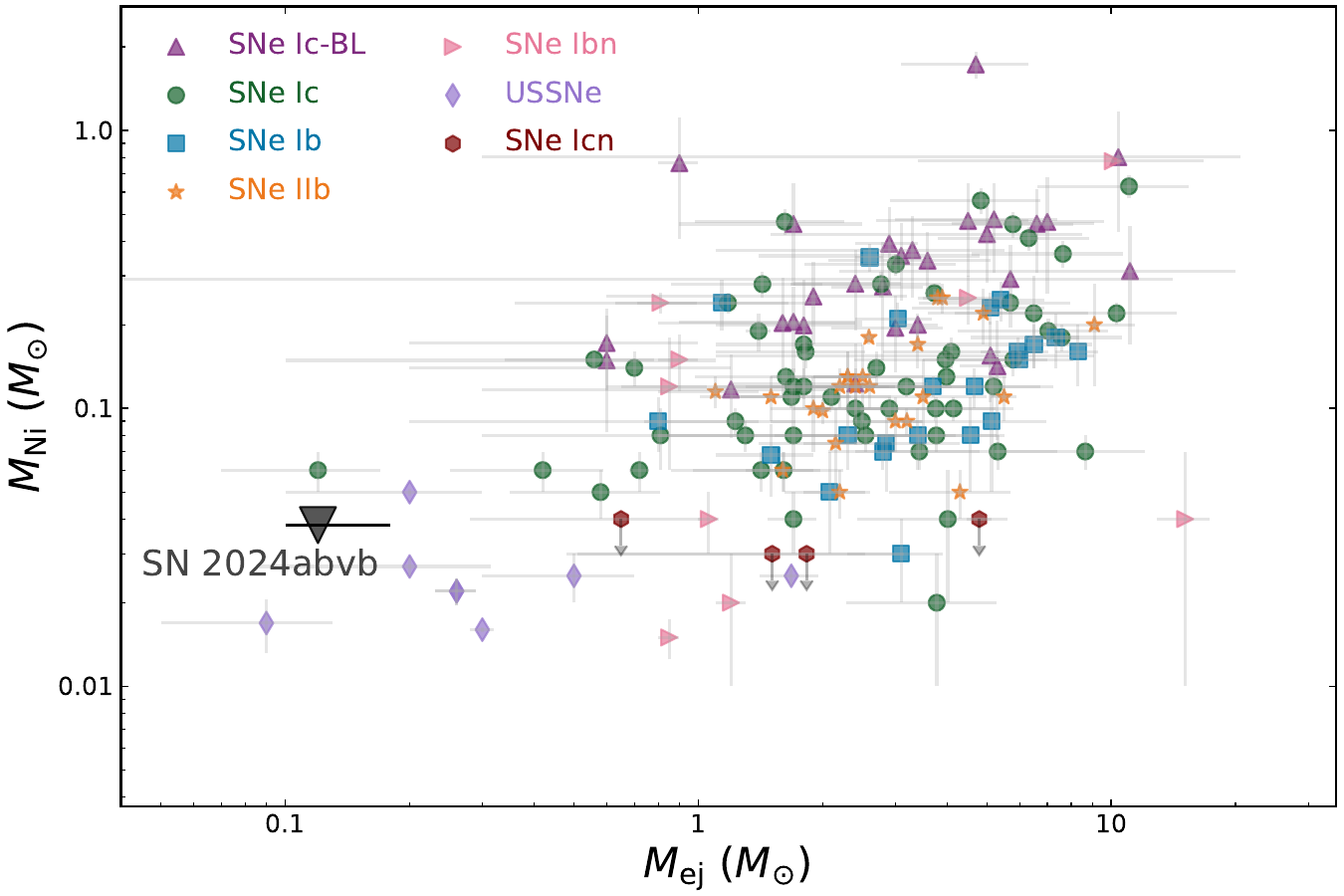}
\caption{The upper triangles, circles, squares, stars, right triangles, diamonds, and hexagons respectively represent the $M_{\rm ej}$--$M_{\rm Ni}$ relationship of broad-lined SNe~Ic (SNe~Ic-BL; \citealt{2018A&A...609A.136T,2019A&A...621A..71T}), SNe~Ic (\citealt{2018A&A...609A.136T,2020MNRAS.497.3770G,2021A&A...651A..81B,2022ApJ...924...55G}), SNe~Ib (\citealt{2018A&A...609A.136T,2020MNRAS.497.3770G,2020A&A...634A..21S,2021MNRAS.507.1229P,2025A&A...693A..13S}), SNe~IIb (\citealt{2018A&A...609A.136T,2022MNRAS.513.5540M,2023ApJ...957..100G,2025ApJ...990L..68S}), SNe~Ibn(\citealt{2008MNRAS.389..113P,2020ApJ...889..170G,2021A&A...652A.136K,2023ApJ...946...30B,2024ApJ...977..254D,2025A&A...700A.156W}), USSNe ( \citealt{2013ApJ...774...58D,2023A&A...675A.201A,2023ApJ...959L..32Y}), and SNe~Icn (\citealt{2022ApJ...938...73P,2023MNRAS.523.2530D}). The black lower triangle corresponds to SN~2024abvb, of which $M_{\rm Ni} < 3.8\times10^{-2}\,M_{\odot}$ is derived from the hybrid model that includes the ejecta-CSM interaction and $^{56}$Ni decay. 
\label{fig_99_Mejecta}}
\end{figure*}

\subsection{Modeling with a Hybrid Model}
\label{Model_lc}

Multiple energy sources may contribute to the early emission of core-collapse SNe. These include the prompt release of electromagnetic emission ahead of the shock wave that ejects the stellar envelope (namely the shock-breakout emission), the recombination emission that mostly takes place in the outermost H-rich envelope and manifests as strong Balmer lines, the decay chain of explosion-synthesized radioactive nuclei (in particular $^{56}$Ni$\rightarrow ^{56}$Co$\rightarrow ^{56}$Fe), the conversion of kinetic energy to radiation facilitated by the SN ejecta expanding into the CSM, and the injection of energy from a central engine such as a magnetar as it spins down (e.g., \citealt{1980ApJ...237..541A,1982ApJ...253..785A,1994ApJ...420..268C,2010ApJ...717..245K,2010ApJ...725..904N,2011ApJ...729...61B,2012ApJ...746..121C,2013MNRAS.435.1520M,2016A&A...589A..53N}). For SN~2024abvb, all these aforementioned mechanisms, except H recombination, remain viable candidates. Detecting prominent emission-line features in SN~2024abvb provides direct observational evidence of CSM interaction. The lack of early-time observations hinders a comprehensive modeling of the shock-cooling process. Furthermore, we did not attempt to model the luminosity evolution by introducing any central engine. Therefore, our final hybrid model incorporates the CSM interaction and the radioactive decay of $^{56}$Ni, both implemented through the Modular Supernova Fitting Tool (MOSFiT; \citealt{2018ApJS..236....6G}). The best-fit parameters and the associated 1$\sigma$ uncertainties were derived using a Markov Chain Monte Carlo (MCMC) approach. 

Within the MOSFiT framework, the $^{56}$Ni decay component is parameterized by the ejecta mass ($M_{\rm ej}$), the kinetic energy ($E_{\rm ej}$), and the $^{56}$Ni mass ($M_{\rm Ni}$). Meanwhile, the ejecta-CSM interaction is characterized by $M_{\rm ej}$, $E_{\rm ej}$, and the mass of the CSM ($M_{\rm CSM}$). The MCMC fitting procedure samples a uniform prior distribution for all fitting parameters within the following ranges: $M_{\rm CSM}$ $\in$ (0.01, 30)\,$M_{\odot}$, $M_{\rm ej}$ $\in$ (0.1, 30)\,$M_{\odot}$, $E_{\rm ej}$ $\in$ (0.05, 5)\,$\times10^{51}\,{\rm erg}$, $M_{\rm Ni}$ $\in$ (0.0001, 1.0)\,$M_{\odot}$, and $t_{\rm exp}$ $\in$ ($-20$, $-8$)\,days. In a parallel configuration, we set walk to 150 and iteration to 5000. The MCMC posterior distributions of the derived parameters and the fit to the multiband light curves of SN\,2024abvb are presented in Figures~\ref{fig_55_mcmc} and \ref{fig_66_LCsFit}, respectively. 
In particular, the best fit to the $\lesssim$ 30\,day light curves gives masses of the ejecta, the CSM, and the radioactive nickel of $M_{\rm ej}=0.12_{-0.02}^{+0.06}$,  $M_{\rm CSM}=0.28_{-0.03}^{+0.02}$, and $M_{\rm Ni}=3.54_{-3.31}^{+34.50}\times10^{^{-3}}$\,$M_\odot$, respectively. The derived parameters are consistent with those estimated from other well-studied cases (e.g., \citealt{2022ApJ...927..180P,2022ApJ...938...73P,2023MNRAS.523.2530D}). 

The posterior distribution of $M_{\rm Ni}$ as presented in the bottom row of Figure~\ref{fig_55_mcmc} shows a long tail toward higher masses. The rather poor constraint on the $^{56}$Ni mass is due to the lack of photometry after $\sim 30$\,day, when the relative energy contribution by the $^{56}$Co$\rightarrow ^{56}$Fe progressively increases. As illustrated in the lower-right panel of Figure~\ref{fig_66_LCsFit}, the radiation around peak brightness is mainly produced by the ejecta-CSM interaction compared to that by the $^{56}$Ni decay. As a result, the near-peak photometry itself cannot place stringent constraints on the mass of radioactive nickel synthesized in the ejecta of the strongly interacting SN\,2024abvb. Therefore, we adopt $M_{\rm Ni} = 3.8\times10^{-2}\ M_{\odot}$ as the upper limit for the synthesized $^{56}$Ni mass of SN~2024abvb.

In the model adapted to the MOSFiT framework, we assume a spherically symmetric emitting zone for the radioactive decay of nickel, and the radiation diffusion throughout the SN ejecta is described following the prescriptions of Eq. 9 of \cite{2012ApJ...746..121C}. We do not consider the diffusion process within the unshocked CSM for the emission generated by either the $^{56}$Ni decay or the ejecta-CSM interaction. 
One possible consequence of this diffusion process is that photons produced in the early stages of CSM interaction may become trapped within the unshocked CSM, causing the ``first light'' to be observed later than the actual explosion time. The simplified approach of ignoring photon diffusion may introduce systematic uncertainties in estimating explosion parameters, such as the explosion time \citep{2025ApJ...984...44H}. 
The exact process by which a photon propagates through the SN ejecta and diffuses out of the surrounding CSM requires a more careful treatment of model parameters, including the density profile, the time-varying optical depth, and the geometries of both the ejecta and the CSM. The calculation also demands computationally intensive Monte Carlo radiative-transfer simulations. Detailed modeling of the photometric evolution of SN\,2024abvb is beyond the scope of this paper. 

As shown in the upper-left panel of Figure~\ref{fig_66_LCsFit}, the observed UV-band peak luminosity is lower than the model predictions, which may be due to the extinction along the line of sight. Moreover, the observed time of the UV-band maximum appears slightly later than the predicted time from the MOSFiT model. As discussed above, ignoring photon diffusion in the unshocked CSM may lead to a discrepancy in the UV-band rising light curve between the MOSFiT model and observations. However, a detailed study of how photon diffusion in the unshocked CSM affects multi-band rising light curves is beyond the scope of this paper.

We also note that our model assumes a homogeneous mass-loss wind from the progenitor star. This isotropic CSM content was found in contradiction to the aspherical CSM directly adjacent to the progenitor from polarimetry (e.g.,~\citealt{2017MNRAS.471.4381S,2026arXiv260200389T}). In some cases, the radial density structure of the CSM deviates from a steady, spherically symmetric mass-loss profile, indicating enhanced episodes of mass loss or a series of mass ejections during outbursts of luminous blue variables (e.g., ~\citealp{1994A&A...290..819L, 2008A&A...483L..47T}). Therefore, we consider the estimated CSM mass a plausible upper limit.

\section{Discussion and Conclusions}
\label{SecIV}

In this paper, we present photometry and spectroscopy of SN~2024abvb, which we classify as an SN~Icn based on the identification of a set of relatively narrow (FWHM $\approx 1000$--2000\,km\,s$^{-1}$), singly ionized carbon emission lines. Such a carbon-rich CSM expanding at a fairly high velocity implies a hydrogen- and helium-depleted, carbon-dominant, Wolf-Rayet-star-like wind (WC type). Among the SN~Icn subclass, the light-curve morphology exhibits significant diversity with an $r$-band peak magnitude ranging from $-$17\,mag (SN\,2019jc, ~\citealp{2022ApJ...938...73P}) to $-$20\,mag (SN\,2021csp, \citealp{2022ApJ...927..180P, 2022ApJ...938...73P}). SN~2024abvb shows a peak brightness similar to that of SN~2021csp but declines faster. 

We fit the UV-to-optical light curves of SN~2024abvb, adopting a hybrid analytic model that accounts for both the ejecta-CSM interaction and the radioactive decay of $^{56}$Ni. The best-fit result suggests a rather small nickel mass of $M_{\rm Ni}\le3.8\times10^{^{-2}}$\,M$_{\odot}$ in the ejecta, and the near-peak emission of SN\,2024abvb is dominated by the ejecta-CSM interaction. The derived explosion time, kinetic energy, $^{56}$Ni mass, and CSM mass are consistent with the results in \cite{2026arXiv260220775A}, although they obtained a relatively larger ejecta mass ($M_{\rm ej} = 3.9\ M_{\odot}$). This difference may arise from differing assumptions about the CSM: we adopted a steady-wind CSM model, whereas \cite{2026arXiv260220775A} used a shell-like CSM model. Besides, \cite{2026arXiv260216227S} fitted the multi-band light curve of SN~2024abvb to constrain the $^{56}$Ni mass of $M_{\rm Ni} \sim 1.0\ M_{\odot}$, indicating the potential $^{56}$Ni-driven picture. The relatively large kinetic energy ($E_{\rm ej} \sim 2.3\times10^{52}\ {\rm erg}$) in \cite{2026arXiv260216227S} produces an expansion velocity of about $3\times10^4\ {\rm km}\,{\rm s}^{-1}$ ($\sim \sqrt{2E_{\rm ej}/M_{\rm ej}}$), possibly resulting in the fast-evolving light curve as SN~2024abvb does. However, the lack of a late-phase light curve for SN~2024abvb hinders the analysis of the synthesized $^{56}$Ni, as the energy contribution from $^{56}$Co $\rightarrow$ $^{56}$Fe increases and the contribution from CSM interaction decreases.

The nature of SNe~Icn still remains controversial. Similar to SNe~Ibn, they may originate from the explosion of WR stars, since both exhibit signatures of CSM that expand at rather high velocities of $\sim 10^{3}\,{\rm km}\,{\rm s}^{-1}$ and have stripped their outer H/He envelopes. The mechanism driving such intense ejection of the outer H/He envelopes immediately prior to the terminal explosion remains unexplained. The absence of apparent P~Cygni profiles in our early-time spectral sequence of SN\,2024abvb hinders the measurement of the wind velocity. The modeled $M_{\rm ej}$ ($\sim 0.12\,M_{\odot}$) and $M_{\rm Ni}$ ($\le 3.8\times10^{-2}\,M_{\odot}$) are smaller than those of stripped-envelope SNe as shown in Figure~\ref{fig_99_Mejecta}, indicating that the progenitor star of SN~2024abvb experienced significant mass loss prior to the explosion. 

In the $M_{\rm ej}$--$M_{\rm Ni}$ relationship shown in Figure~\ref{fig_99_Mejecta}, it seems that SN~2024abvb might be consistent with the ultrastripped-envelope SN (USSN) group, indicating that they may share a similar physical origin. The key difference lies in the presence of dense confined CSM around SN~2024abvb, whereas USSNe exhibit a relatively clean circumstellar environment. This distinction accounts for the higher luminosity of SN~2024abvb compared to USSNe (e.g., $r$-band peak magnitude $\sim -16.2$ for SN~2021agco \citealt{2023ApJ...959L..32Y}). If the underlying explosions are indeed the same, SN~2024abvb should exhibit nebular-phase spectral features comparable to those of USSNe. With the increasing number of time-domain surveys, the physical origins linking SNe~Icn to USSNe will be further constrained.

SN~2024abvb exploded in the outskirts of its host galaxy, which may correspond to a relatively long evolving history for its progenitor system. One plausible explanation for the small ejecta mass and nickel mass associated with the bright, fast-declining light curve of SN~2024abvb is that the exploding progenitor star originated in a binary system consisting of a helium star and a compact companion \citep{2015MNRAS.451.2123T}. The H and He envelopes of the former are progressively stripped by the latter through Roche-lobe overflow \citep{2007ASPC..372..397M}. Such enhanced mass loss toward the orbital plane of the binary system would take place when the Roche lobe is filled, and is suggested to be capable of efficiently stripping the H envelope~\citep{2014ARA&A..52..487S}. \cite{2026arXiv260200389T} reported the time-variant polarimetric signals and the presence of Balmer absorptions and He lines in their high-resolution spectra. These observations suggest that SN~2024abvb arises from a binary system undergoing multiple mass-loss episodes with different orientations during the common-envelope phase. While this scenario adds further details, it fundamentally aligns with the physical picture described above.

The chemical composition and wind velocity of the CSM encode information about the mass-loss history of massive stars nearing the end of their lives. The fusion reactions transform massive stars into a structure of concentric shells. One of the key deliverables of stripped-envelope SNe is testing the long-hypothesized internal layering that develops at the end of the evolution of massive stars: progressively heavier elements are synthesized toward the iron core at the center. Spectrophotometric observations from the earliest possible phases would reveal the composition and physical conditions of the surface layers of the exploding star, which are immediately expelled before the explosion. Observations of Type Ibn/Icn SNe, together with the very recently reported Type Ien SN\,2021yfj \citep{2024TNSAN.239....1G, 2024TNSAN.240....1G, 2024arXiv240902054S}, whose immediate CSM exhibits chemical signatures of silicon, sulfur, and argon, provide key clues for understanding the nature and process of such a stripping sequence emerging from the death of massive stars.

\begin{acknowledgments}
We thank Yi Yang for helpful discussions.
This work is supported by the National Natural Science Foundation of China (NSFC grants 12288102, 12403049, 12033003, and 12373038), China Postdoctoral Science Foundation (Certificate Number 2025T180872), Natural Science Foundation of Xinjiang Uygur Autonomous Region under No. 2024D01D32, and Tianshan Talent Training Program grant 2023TSYCLJ0053. 
This work is partly supported by the Urumqi Nanshan Astronomy and Deep Space Exploration Observation and Research Station of Xinjiang (XJYWZ2303).
Maokai Hu acknowledges support from the Postdoctoral Fellowship Program of CPSF under grant GZB20240376, and the Shuimu Tsinghua Scholar Program 2024SM118. 
X. Wang acknowledges support from the Tencent Xplorer Prize.
Jujia Zhang is supported by the B-type Strategic Priority Program of the Chinese Academy of Sciences (grant XDB1160202), the National Key R\&D Program of China with grant 2021YFA1600404, NSFC grants 12173082 and 12333008, the Yunnan Fundamental Research Projects (YFRP grants 202501AV070012 and 202401BC070007), the Top-notch Young Talents Program of Yunnan Province, and the International Centre of Supernovae, Yunnan Key Laboratory (grant 202302AN360001). 
Lingzhi Wang is sponsored by NSFC grant 12573050, the Chinese Academy of Sciences South America Center for Astronomy (CASSACA) Key Research Project E52H540301, and in part by the Chinese Academy of Sciences (CAS) through a grant to the CASSACA.
A.V.F.’s research group at UC Berkeley acknowledges financial                 
assistance from the Christopher R. Redlich Fund, as well as donations         
from William Draper, Timothy and Melissa Draper, Briggs and Kathleen Wood, Sanford Robertson (T.G.B. is a Draper-Wood-Robertson Specialist in Astronomy),
and numerous other donors.
A.V.F. is grateful for the hospitality of the Hagler Institute for Advanced Study as well as the Department of Physics and Astronomy at Texas A\&M University during part of this investigation.

We acknowledge the support of the staff of the Xinglong 2.16\,m telescope, the Lijiang 2.4\,m telescope, the 3\,m Lick/Shane telescope, the REM telescope, and the Tsinghua-Nanshan Optical Telescope.  Funding for the LJT has been provided by the CAS and the People's Government of Yunnan Province. The LJT is jointly operated and administered by YNAO and the Center for Astronomical Mega-Science, CAS.
A major upgrade of the Kast spectrograph on the Shane 3\,m telescope at        
Lick Observatory, led by Brad Holden, was made possible through    
generous gifts from the Heising-Simons Foundation, William and Marina         
Kast, and the University of California Observatories.  Research at            
Lick Observatory is partially supported by a generous gift from Google.

\end{acknowledgments}

\begin{table*}[h]
    \centering
    \begin{tabular}{ccccccc}
    \hline
   MJD  &   Phase & Filter  & Magnitude  & Instrument & System  \\ 
    \hline
60642.601 & -2.149    &$B$ & 17.10$\pm$0.10 & TNOT  &  Vega   \\
60642.602 & -2.148   & $B$ & 17.10$\pm$0.08 & TNOT  &  Vega   \\
60642.604 & -2.146   & $B$ & 17.07$\pm$0.06 & TNOT  &  Vega   \\
60642.605 & -2.145   & $B$ & 17.09$\pm$0.08 & TNOT  &  Vega   \\
60643.577 & -1.174   & $B$ & 17.03$\pm$0.10 & TNOT  &  Vega   \\
60643.578 & -1.172   & $B$ & 17.03$\pm$0.13 & TNOT  &  Vega   \\
60643.579 & -1.171   & $B$ & 17.05$\pm$0.06 & TNOT  &  Vega   \\
60643.580 & -1.170   & $B$ & 17.05$\pm$0.11 & TNOT  &  Vega   \\
60644.673 & -0.077   & $B$ & 17.02$\pm$0.13 & TNOT  &  Vega   \\
60644.675 & -0.075   & $B$ & 17.01$\pm$0.08 & TNOT  &  Vega   \\
60644.676 & -0.074   & $B$ & 17.02$\pm$0.07 & TNOT  &  Vega   \\
60644.677 & -0.073   & $B$ & 17.02$\pm$0.07 & TNOT  &  Vega   \\
60649.611 & 4.861   & $B$ & 17.35$\pm$0.18 & TNOT  &  Vega   \\
60650.646 & 5.896   & $B$ & 17.27$\pm$0.14 & TNOT  &  Vega   \\
60651.557 & 6.808   & $B$ & 17.49$\pm$0.08 & TNOT  &  Vega  \\
60652.599 & 7.849   & $B$ & 17.61$\pm$0.12 & TNOT   &  Vega  \\

... ... &  &   &   &   &  \\
    \hline
    \end{tabular}
    \caption{Photometry of SN~2024abvb. The magnitudes are not corrected for extinction in the host galaxy or the Milky Way, and the phase is relative to the time of $r$-band peak brightness. The complete machine-readable table is available online. }
    \label{Table_33_LCs}
\end{table*}

\begin{table*}[ht!]
    \centering
    \begin{tabular}{cccccccc}
    \hline
   No.  &  UTC Date
& MJD &  Phase (d)  & Exp. (s)  & Telescope+Inst.  & Range (\AA) & Airmass  \\ 
    \hline
   1     &    11.27 2024    &    60641.90    &  $-2.9$    & 1800     &   NOT+ALFOSC   &  3700-8000   &  /      \\ 
   2     &    11.29 2024    &    60643.59    &  $-1.2$    & 1800     &  LJT+YFOSC   &  3600-8000    & 1.2       \\ 
   3     &    11.30 2024    &    60644.61    &  $-0.1$    & 3000     &  XLT+BFOSC   &  4000-8000    & 2.1    \\ 
   4     &    12.01 2024    &    60645.65    &   0.9   & 1800     & LJT+YFOSC    &  3600-8000     & 1.4    \\ 
   5     &    12.02 2024    &    60646.51    &  1.8     & 3600     &   XLT+BFOSC  &  4000-8000    & 1.5     \\ 
   6     &    12.03 2024    &    60647.66    &   2.9    & 3600     &  XLT+BFOSC   &  4000-8000    & 3.6     \\ 
   7     &    12.04 2024    &    60648.49    &  3.7    &  3600    &   XLT+BFOSC  &   3750-8000    & 1.4    \\ 
   8     &    12.05 2024    &   60649.68     &  4.9     & 1700     &   LJT+YFOSC  &  3600-8000    & 1.7     \\ 
   9     &    12.08 2024    &   60652.68     &  7.9    &  1500     &   LJT+YFOSC  &  3600-8000    & 1.8     \\ 
   10   &     12.10 2024    &   60654.23     &  9.5    &  3600     &         Shane+Kast &  4000-8000   &  1.5    \\
   11    &    12.18 2024    &   60662.55     &  17.8   &  2400     &  LJT+YFOSC   &  4000-8000      & 1.2   \\ 
    
    \hline
    \end{tabular}
    \caption{Log of spectroscopic observations of SN~2024abvb. 
    The spectrum observed by NOT is taken from the Transient Name Server. All phases are given relative to the time of $r$-band maximum brightness at MJD\,60644.75. }
    \label{Table_22_SpecInfor}
\end{table*}


\bibliography{sample7}{}
\bibliographystyle{aasjournalv7}



\end{document}